\documentclass[letterpaper,twocolumn,prl,aps,superscriptaddress,amsmath,amssymb,floatfix]{revtex4-1}
\usepackage{mathptmx}
\usepackage[latin9]{inputenc}
\usepackage{color}
\usepackage{amsmath}
\usepackage{amssymb}
\usepackage{graphicx}
\usepackage{esint}
\usepackage[unicode=true,
 bookmarks=true,bookmarksnumbered=false,bookmarksopen=false,
 breaklinks=false,pdfborder={0 0 1},backref=false,colorlinks=true]
 {hyperref}
\hypersetup{
 linkcolor=magenta,urlcolor=blue,citecolor=blue,pdfstartview={FitH},hyperfootnotes=false}

\makeatletter

\pdfpageheight\paperheight
\pdfpagewidth\paperwidth

\usepackage{textcomp}
\usepackage{epstopdf}

\usepackage{amsfonts}

\pdfpageheight\paperheight
\pdfpagewidth\paperwidth

\@ifundefined{textcolor}{}{%
 \definecolor{BLACK}{gray}{0}
 \definecolor{WHITE}{gray}{1}
 \definecolor{RED}{rgb}{1,0,0}
 \definecolor{GREEN}{rgb}{0,1,0}
 \definecolor{BLUE}{rgb}{0,0,1}
 \definecolor{CYAN}{cmyk}{1,0,0,0}
 \definecolor{MAGENTA}{cmyk}{0,1,0,0}
 \definecolor{YELLOW}{cmyk}{0,0,1,0}
}

\usepackage{xcolor}\usepackage{soul}
\definecolor{blue}{rgb}{0,0,1}
\definecolor{red}{rgb}{1,0,0}
\definecolor{green}{rgb}{0,1,0}

\makeatother

\begin{document}
\title{Photon-photon quantum phase gate in a photonic molecule with $\chi^{(2)}$
nonlinearity}
\author{Ming~Li}
\affiliation{Key Laboratory of Quantum Information, Chinese Academy of Sciences,
University of Science and Technology of China, Hefei 230026, P. R. China.}
\affiliation{CAS Center For Excellence in Quantum Information and Quantum Physics,
University of Science and Technology of China, Hefei, Anhui 230026,
P. R. China.}
\author{Yan-Lei~Zhang}
\affiliation{Key Laboratory of Quantum Information, Chinese Academy of Sciences,
University of Science and Technology of China, Hefei 230026, P. R. China.}
\affiliation{CAS Center For Excellence in Quantum Information and Quantum Physics,
University of Science and Technology of China, Hefei, Anhui 230026,
P. R. China.}
\author{Hong X.~Tang}
\affiliation{Department of Electrical Engineering, Yale University, New Haven,
CT 06511, USA}
\author{Chun-Hua~Dong}
\email{chunhua@ustc.edu.cn}

\affiliation{Key Laboratory of Quantum Information, Chinese Academy of Sciences,
University of Science and Technology of China, Hefei 230026, P. R. China.}
\affiliation{CAS Center For Excellence in Quantum Information and Quantum Physics,
University of Science and Technology of China, Hefei, Anhui 230026,
P. R. China.}
\author{Guang-Can~Guo}
\affiliation{Key Laboratory of Quantum Information, Chinese Academy of Sciences,
University of Science and Technology of China, Hefei 230026, P. R. China.}
\affiliation{CAS Center For Excellence in Quantum Information and Quantum Physics,
University of Science and Technology of China, Hefei, Anhui 230026,
P. R. China.}
\author{Chang-Ling~Zou}
\email{clzou321@ustc.edu.cn}

\affiliation{Key Laboratory of Quantum Information, Chinese Academy of Sciences,
University of Science and Technology of China, Hefei 230026, P. R. China.}
\affiliation{CAS Center For Excellence in Quantum Information and Quantum Physics,
University of Science and Technology of China, Hefei, Anhui 230026,
P. R. China.}
\affiliation{National Laboratory of Solid State Microstructures, Nanjing University,
Nanjing 210093, China.}
\date{\today}
\begin{abstract}
The construction of photon-photon quantum phase gate based on photonic
nonlinearity has long been a fundamental issue, which is vital for
deterministic and scalable photonic quantum information processing.
It requires not only strong nonlinear interaction at the single-photon
level, but also suppressed phase noise and spectral entanglement for
high gate fidelity. In this paper, we propose that high-quality factor
microcavity with strong $\chi^{(2)}$ nonlinearity can be quantized
to anharmonic energy levels and be effectively treated as an artificial
atom. Such artificial atom has a size much larger than the photon
wavelength, which enables passive and active ultra-strong coupling
to traveling photons. High-fidelity quantum control-phase gate is
realized by mediating the phase between photons with an intermediate
artificial atom in a photonic molecule structure. The scheme avoids
the two-photon emission and thus eliminates the spectral entanglement
and quantum phase noises. Experimental realization of the artificial
atom can be envisioned on the integrated photonic chip and holds great
potential for single-emitter-free, room-temperature quantum information
processing.
\end{abstract}
\maketitle
\emph{Introduction.- }Quantum photonic integrated circuit (PIC) has
been extensively studied since the last decade for photon-based quantum
information processing$\;$\cite{PolitiNov.,Takeda2019,Ladd2010,quantumsimulator,shoragr},
due to its advantages of stability, compactness and low power consumption.
Essential quantum optical components, including quantum photon sources$\;$\cite{photonsourceReview},
quantum gates$\;$\cite{Politi2008,Crespi2011} and single-photon
detectors$\;$\cite{Najafi2015,Cheng2016,sspd2019}, have been all
demonstrated on the PIC with excellent performances, and fully integrated
quantum PIC is within reach$\;$\cite{Fullcircuit2016}. However,
the absence of single-photon nonlinearity greatly limits the development
of PIC for scalable quantum processors$\;$\cite{boyd2003nonlinear},
since the deterministic quantum gates among photons are forbidden.
For example, a photon-photon quantum phase gate requires the controlling
of the phase of one photon by another photon$\;$\cite{Nielsen2010}.
It is believed that the intrinsic nonlinear effect of a dielectric
is too weak compared with the material absorption$\;$\cite{boyd2003nonlinear},
thus the desired quantum operation can not be accomplished before
the photon is lost. As a result, the photonic two-qubit quantum gates
are implemented probabilistically with pure linear optical components
and rely on quantum interference and ancillary photons$\;$\cite{Politi2008,OBrien2009}.
Another approach to overcome this obstacle is introducing single emitters
into the PIC, while suffering from the experimental challenges of
nano-manipulation and instabilities of emitters$\;$\cite{photon-photon-rydberg,Tiecke2014,Hacker2016}.

Fortunately, recent exciting progress in nonlinear optics on PIC has
encouraged the efforts to pursue nonlinearity at the single-photon
level, with the help of sophisticated fabrication technique, new material,
and advanced photonic structure engineering. By realizing the micro-
and nano-resonators of ultrahigh quality factor and ultrasmall mode
volume, the cavity photonic nonlinear interaction strength is greatly
boosted while the losses are suppressed$\;$\cite{vahala2003optical,strekalov2016nonlinear,Lin:17,Bruch2019}.
In the past few years, cavity-enhanced nonlinear photonics has achieved
great success in frequency conversion, frequency comb, and quantum
photon sources$\;$ \cite{li2016efficient,Guo2016a,strekalov2016nonlinear,Kippenberg2018,li2018optimal}.
Especially, ultrahigh-efficiency second-harmonic generation (SHG)
with efficiency as high as $10^{3}-10^{6}\,\mathrm{\%/W}$ are achieved
\cite{Bruch17000,chang2019strong,chengya,LN2019}. All these exciting
progresses achieved with $\chi^{(2)}$, such as in lithium niobate
(LN)$\;$\cite{zhang2017monolithic,Zhang2019,LN2019,ChenLN2019},
aluminum nitride (AlN)$\;$\cite{Guo2016,Bruch17000} and gallium
arsenide (GaAs)$\;$\cite{chang2019strong} indicate a saturation
of conversion efficiency even at single-photon level pump, and reveal
a promising path towards single-photon nonlinearity.

However, there is still another obstacle for the deterministic photonic
quantum gates, since they demand the processing of photon's quantum
states while maintaining their spectral-temporal wavefunction. As
pointed out by Shapiro$\;$\cite{PhysRevA.73.062305,PhysRevA.81.043823,PhysRevA.90.062314},
the fidelity of photon-photon quantum gate based on photonic nonlinearity,
such as Kerr effect, suffers from spectral entanglement and phase
noise, due to the spatially-local interaction and multimode nature
of traveling photons in the frequency domain. Even though several
schemes have been proposed to overcome such limitation by introducing
non-local interaction and cascaded sites$\;$\cite{PhysRevA.87.042325,PRL2016MultiSite,xia2016},
these works are still based on propagation modes and the optical loss
in nonlinear media is not considered. Besides, the Kerr nonlinearity
is mediated by neutral atoms, which has finite linewidth and its integration
with the photonic chip is challenging in practice.

In this work, we propose an artificial atom on PIC by utilizing $\chi^{(2)}$
nonlinearity in a well-engineered microresonator. The artificial atom
has a size of microns, and is thus easy for fabrication and is scalable.
Compared with neutral atoms, the artificial atoms can strongly couple
with a waveguide, so the photon could be stored and extracted efficiently.
Additionally, the artificial atom possesses degenerate chiral energy
levels, so allowing for mediating unidirectional photon-photon interactions.
As an example, we proposed an architecture for realizing quantum control-$Z$
($CZ$) gate based on a photonic molecule$\;$\cite{Zhang2019}. By
treating one of the resonators as a tunable antenna for coupling the
photons with the other artificial atom, the single photons can be
stored and coupled with each other strongly, breaking the limitations
of the spatially local interaction condition for traveling photons,
thus being immune from spectral entanglement and phase noises. With
potentially achievable parameters in the experiment, we predict a
fidelity of $CZ$ gate of $99\%$. Combining with the mature single-qubit
gates (Hadamard and phase gates) by linear optical elements, the universal
gate sets~\cite{Nielsen2010} could be accomplished and the scalable
quantum computation is promising on the PIC platform.

\begin{figure}
\begin{centering}
\includegraphics[width=0.95\columnwidth]{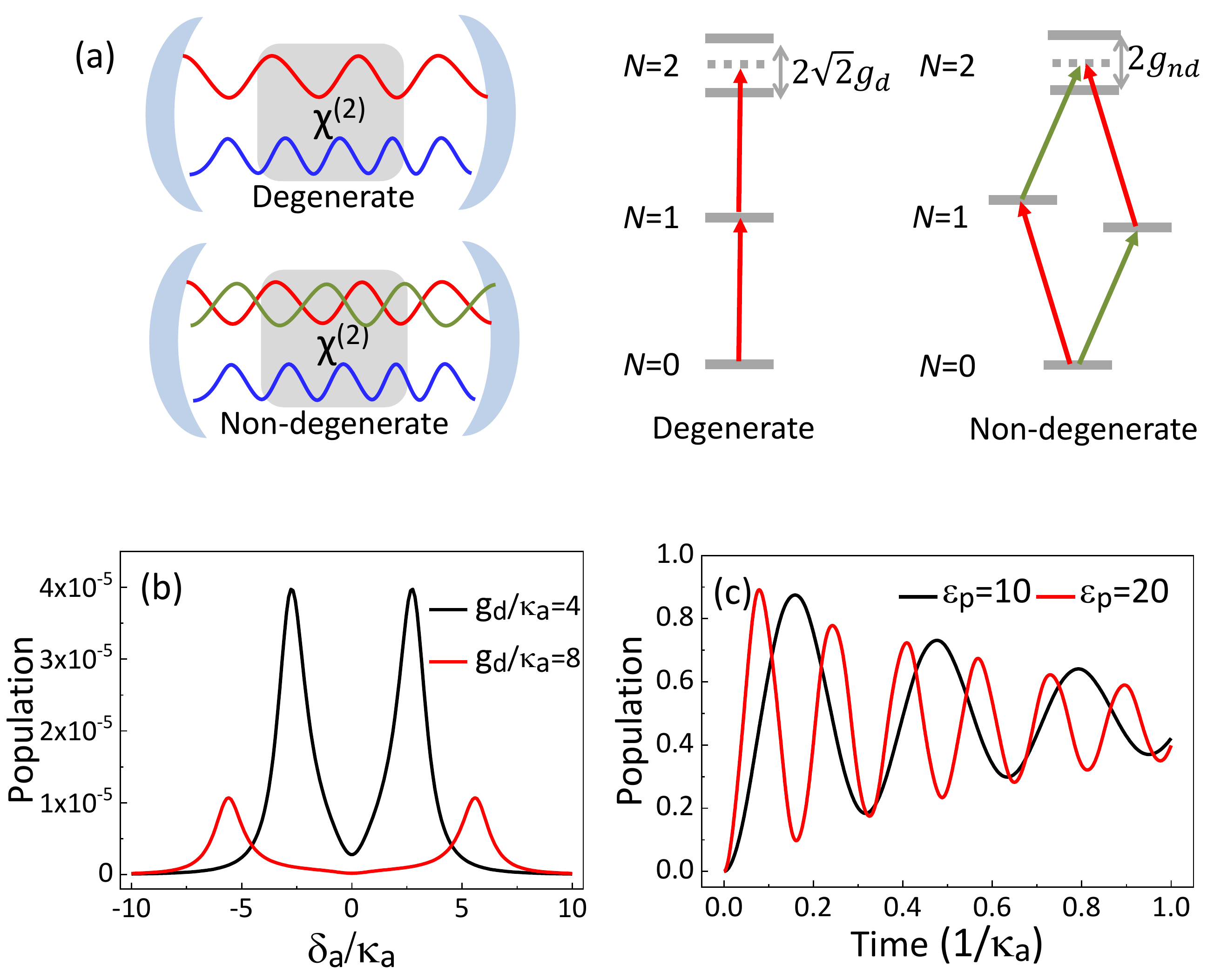}
\par\end{centering}
\caption{Artificial atom based on an optical cavity with strong $\chi^{(2)}$
nonlinearity. (a) Left: schematic of degenerate and non-degenerate
three-wave mixing, corresponding to second-harmonic generation (SHG)
and sum-frequency generation (SFG). Right: energy level structures
for $\chi^{(2)}$ interaction. The energy levels of two-excitation
state split by $2\sqrt{2}g_{\mathrm{d}}$ for degenerate case and
$2g_{\mathrm{nd}}$ for non-degenerate case due to $\chi^{(2)}$ interaction.
(b) Vacuum Rabi splitting of the populations of state $|2_{a}0_{c}\rangle$
for SHG in the cavity. The coupling strengths are set $g_{\mathrm{d}}/\kappa_{a}=4$
for the black curve and $g_{\mathrm{d}}/\kappa_{\mathrm{a}}=8$ for
the red curve. The driving strength is $\varepsilon_{p}=0.2$. The
Rabi splitting is proportional to the nonlinear coupling strength.
(c) Rabi oscillation of the single-photon excitation state $|1_{a}0_{c}\rangle$
population under coherent driving. The coupling strength is $g_{\mathrm{d}}/\kappa_{a}=80$
and the driving strength is $\varepsilon_{p}=10$ for the black curve
$\varepsilon_{p}=20$ for the red curve. The cavity decay rates of
the fundamental and second-harmonic modes are set $\kappa_{\mathrm{a}}=\kappa_{\mathrm{c}}=1$.}

\label{Fig1}
\end{figure}

\emph{Artificial atom.- }Figure$\,$\ref{Fig1}(a) schematically illustrates
the artificial atom, which is based on phase-matched $\chi^{(2)}$
ultrahigh-Q microresonator. In the view of nonlinearly coupled optical
modes, the system Hamiltonian reads~\cite{Guo2016a,Bruch17000} $\left(\hbar=1\right)$
\begin{equation}
H_{\mathrm{AA}}=\sum_{j}\omega_{j}j^{\dagger}j+g_{\mathrm{d}}\left(a^{\dagger2}c+a^{2}c^{\dagger}\right),
\end{equation}
with $j\in\{\mathrm{a},\mathrm{c}\}$ and the mode frequency $2\omega_{\mathrm{a}}\approx\omega_{\mathrm{c}}$
for degenerate three-wave mixing process. For the non-degenerate case
$j\in\{\mathrm{a},\mathrm{b},\mathrm{c}\}$, the interaction reads
$g_{\mathrm{nd}}\left(a^{\dagger}b^{\dagger}c+abc^{\dagger}\right)$
with $\omega_{\mathrm{a}}\neq\omega_{\mathrm{b}}$ and $\omega_{\mathrm{b}}\approx\omega_{\mathrm{c}}-\omega_{\mathrm{a}}$.
Here, $a,b,c$ denotes the bosonic operator for the modes, and $g_{\mathrm{d},\mathrm{nd}}\propto\chi^{(2)}\xi/\sqrt{V}$
is the coupling strength which is determined by the modal overlap
$\xi$, cavity mode volume $V$, and the material nonlinear susceptibility
$\chi^{(2)}$.

In conventional bulk nonlinear optics, the coupling rate is much weaker
than the dissipation rate ($g\ll\kappa$) due to large $V$, thus
the conversion from $a,b$ to $c$ only occurs by strongly pumping
the system~\cite{Guo2016a,Bruch17000}. With improved $\xi$ and
drastically reduced $V$ in PIC, $g$ increases and approaches $\kappa$,
and thereby the quantum effect appears, as shown by the right panel
of Fig.$\,$\ref{Fig1}(a). Considering only few excitations in those
modes, the system energy levels can be rewritten as $\left|l_{\mathrm{a}}n_{\mathrm{c}}\right\rangle $
or $\left|l_{\mathrm{a}}m_{\mathrm{b}}n_{\mathrm{c}}\right\rangle $
by Fock state basis, with $l,m,n\in\mathbb{Z}$ (the subscripts of
the states are omitted in the following). Due to the nonlinear interaction
in degenerate case, the energy levels $\left|20\right\rangle $ and
$\left|01\right\rangle $ are hybridized and produce eigenstates $\left(\left|20\right\rangle \pm\left|01\right\rangle \right)/\sqrt{2}$,
whose energy levels shifted by $\pm\sqrt{2}g_{\mathrm{d}}$, respectively.
Similarly, for the non-degenerate case, the state $\left|110\right\rangle $
strongly couples with $\left|001\right\rangle $ and produces new
eigenstates $\left(\left|110\right\rangle \pm\left|001\right\rangle \right)/\sqrt{2}$
with a frequency splitting $2g_{\mathrm{nd}}$. Then, the cavity shows
anharmonicity for Fock states of modes $a$ (or $a$ and $b$), as
shown by the dashed lines in Fig.$\,$\ref{Fig1}(a). When $\kappa<g_{\mathrm{d},\mathrm{nd}}$,
the second photon can not enter into the cavity if mode $a$ is already
occupied by one single-photon, which would lead to the photon-blockade
effect$\;$\cite{KerrBlock,chi2blcok,AtomBlock}. In this case, we
treat the cavity as an artificial atom. It is worth noting that, the
artificial atom based on the anharmonicity of Fock states shares the
same spirit of superconducting qubit$\;$\cite{Krantz2019}, which
is an LC circuit with strong anharmonicity and is usually treated
as a two-level system.

To verify the artificial atom, a weak coherent driving $\varepsilon_{\mathrm{p}}\left(a^{\dagger}+a\right)$
is used to probe the system with $g_{\mathrm{d}}/\kappa_{\mathrm{a}}=4$
and $g_{\mathrm{d}}/\kappa_{\mathrm{a}}=4$ for the degenerate case.
Figure.$\,$\ref{Fig1}(b) shows the dependence of the $|20\rangle$
state population on the frequency of a continuous driving field ($\varepsilon_{\mathrm{p}}=0.2$).
The population at zero detunings is greatly suppressed due to the
splitting of the hybrid energy levels. Furthermore, the temporal behavior
of the artificial atom under the on-resonance driving shows clear
Rabi oscillation {[}Fig.$\,$\ref{Fig1}(c){]}, since only $|00\rangle$
and $|10\rangle$ can be effectively excited, thus confirms the equivalence
of our system to a two-level atom. The blockade effect to the high-excitation
energy levels can be used to build a deterministic single-photon source.

\begin{figure}
\begin{centering}
\includegraphics[width=1\columnwidth]{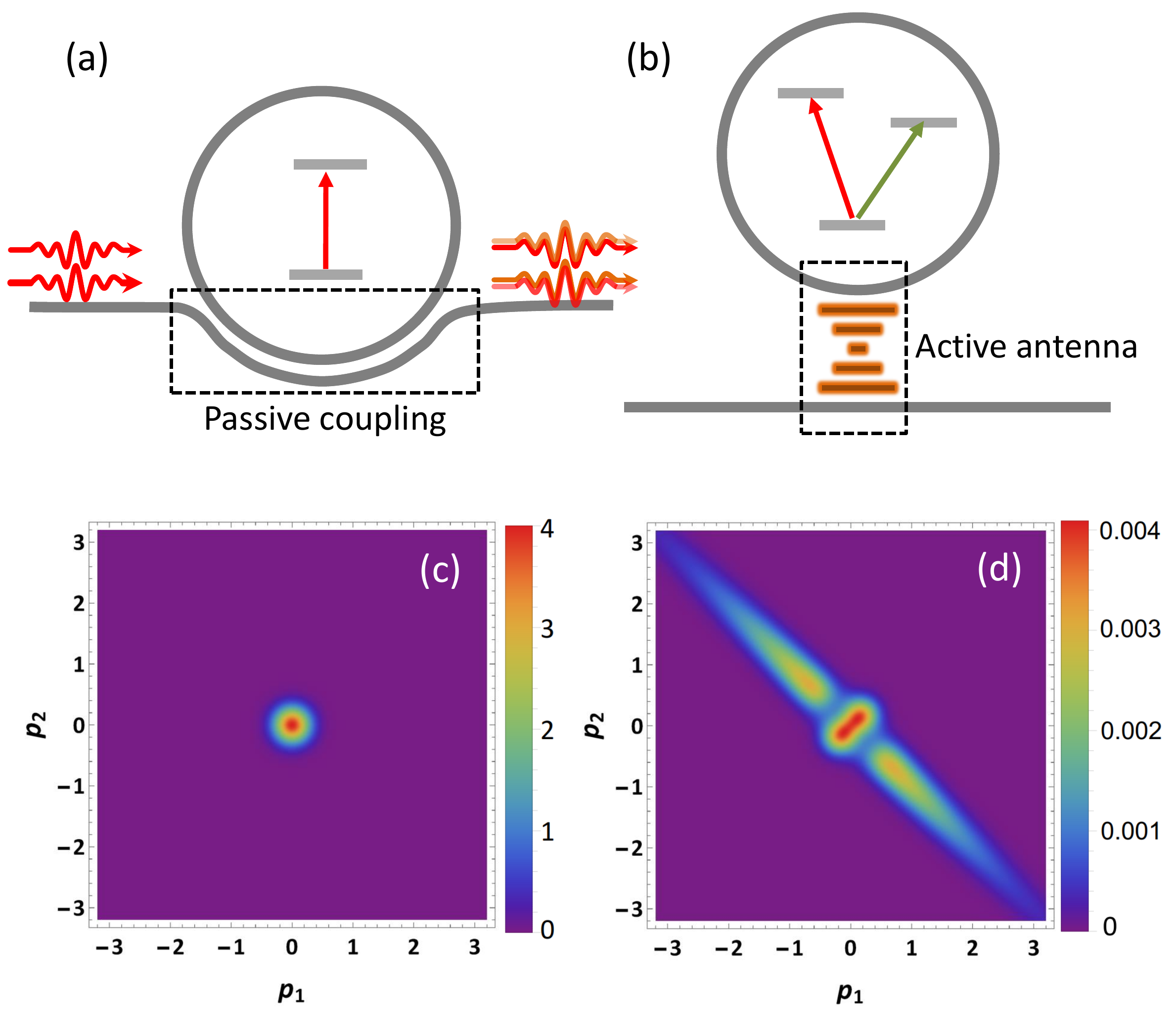}
\par\end{centering}
\caption{Passive and active coupling between waveguide and artificial atom
made by microring cavity. (a)-(b) The photonic designs for the coupling
between waveguide and artificial atom. (a) The artificial atom is
coupled with a waveguide of wrapped-around structure. The coupling
rate can be three orders of magnitude than the decay rate. (b) The
atom-waveguide coupling is dynamically controlled by an active antenna.
Photons of different pulse shapes and frequencies can be coupled to
different energy levels of the artificial atom. (c)-(d) Joint spectral
density of two-photon state. (c) Input state of separable two-photon
state with Gaussian shape $g\left(p_{1},p_{2}\right)=\frac{1}{2\pi\sigma^{2}}\exp\left[-\left(p_{1}^{2}+p_{2}^{2}\right)/2\sigma^{2}\right]$,
with $p_{1,2}$ is the frequency of the photon and $\sigma$ is the
width of the momentum distribution. (d) Output state. In the calculation,
the nonlinear coupling strength $g_{\mathrm{d}}=3$, the decay rates
of the second-harmonic mode $\kappa_{\mathrm{c},0}=\kappa_{\mathrm{c},1}=1$
and the fundamental mode $\kappa_{\mathrm{a},0}=1$, $\kappa_{\mathrm{a},1}=3$,
where $\kappa_{\mathrm{a}(\mathrm{c}),0}$ is the intrinsic decay
rate and $\kappa_{\mathrm{a}(\mathrm{c}),1}$ is the external coupling
rate to the waveguide. The spectral width of the input pulse $\sigma=0.5$.
The output two-photon state shows strong spectral correlation.}

\label{Fig2}
\end{figure}

Compared with natural atoms, the artificial atom maintains many advantages
of photonic cavities. Firstly, degenerate clockwise (CW) and counter-clockwise
(CCW) modes are supported in a traveling wave microresonator. Therefore,
the artificial atom can couple with external photons unidirectionally
{[}Fig.$\,$\ref{Fig2}(a){]}, which is only possible for natural
atoms with photonic spin-orbital coupling$\;$\cite{Lodahl2017}.
Especially, the two-fold degeneracy of modes enables the construction
of two identical artificial atoms with only one design, offering more
energy levels for quantum information processing. Secondly, the artificial
atoms are more flexible for photonic structure designs and allow highly
efficient coupling with a waveguide or other photonic structures.
For studying waveguide quantum electrodynamics~\cite{Zheng2013},
the artificial atom to waveguide coupling can achieve a Purcell factor
$F_{\mathrm{p}}$ exceeding $1000$ {[}with the configuration in Fig.$\,$\ref{Fig2}(a){]}.
In contrast, the achievable $F_{\mathrm{p}}$ for natural atom-waveguide
coupling is limited to $\mathcal{O}\left(1\right)$. In addition to
the passive design, the interaction between the artificial atom and
photons could also be dynamically controlled by active antennas {[}Fig.$\,$\ref{Fig2}(b){]}.
The energy levels of artificial atoms are also reconfigurable, as
the resonance frequencies can be dynamically tuned by external drive
fields via the electro-optic (EO) or thermal-optic effects.

\emph{CZ gate.- }These merits of artificial atoms make it an excellent
platform for realizing scalable quantum gates for photons or atoms.
Employing the photon blockade effect in strongly coupled artificial
atom and waveguide, the probe single photon would gain a $\pi$ phase
when passing the artificial atom if its frequency is on-resonance
with its transition {[}see the Supporting Materials (SM) for details{]}.
However, if the photon is off-resonance with the transition, the artificial
atom would not induce a phase shift. Therefore, it is anticipated
that one photon could induce a $\pi$-phase shift of another photon,
manifesting the $CZ$ gate for photons. We numerically tested the
artificial atom by coupling it to a bus waveguide {[}Fig.$\,$\ref{Fig2}(a){]}
with a input of two cooperating photons {[}Fig.$\,$\ref{Fig2}(c){]}.
The outcome {[}Fig.$\,$\ref{Fig2}(d){]} shows a broaden spectrum
distribution and undesired frequency correlation between the two photons.
The output is a superposition of the directly transmitted state and
the two-photon bound state$\;$\cite{transport-fan,PhysRevA.92.063817,PhysRevA.91.043845},
which means the frequency of the output photons are entangled. Therefore,
it is failed to obtain the $CZ$ gate because the artificial atom
not only induces a $\pi$-phase change but also induces the entanglement
of other degree of freedom. Such a problem has also been predicted
in previous Kerr-type nonlinear medium$\;$\cite{xia2016,PhysRevA.87.042325},
and the reason should be attributed to the continuum modes in the
waveguide: the excitation states $\left|01\right\rangle $ and $\left|20\right\rangle $
of artificial atom could spontaneously emit photon pairs without conserving
the frequency of individual photons, although the total energy of
two photons conserves.

The direct two-photon interaction mediated by the artificial atom
prevents building a quantum logic gate. In other words, the two-photon
spontaneous emission channels of the artificial atom must be suppressed
during the gate operation, thus the coupling between the continuum
and the energy levels of the artificial atom for either control or
target photon should be shut-off. Therefore, we introduce a scheme
based on storing the control photon in the artificial atom, to avoid
the decay of control photon level to the continuum during its interaction
with the target photon, and realize a photon-photon quantum phase
gate on PIC.

\begin{figure}
\begin{centering}
\includegraphics[width=1\columnwidth]{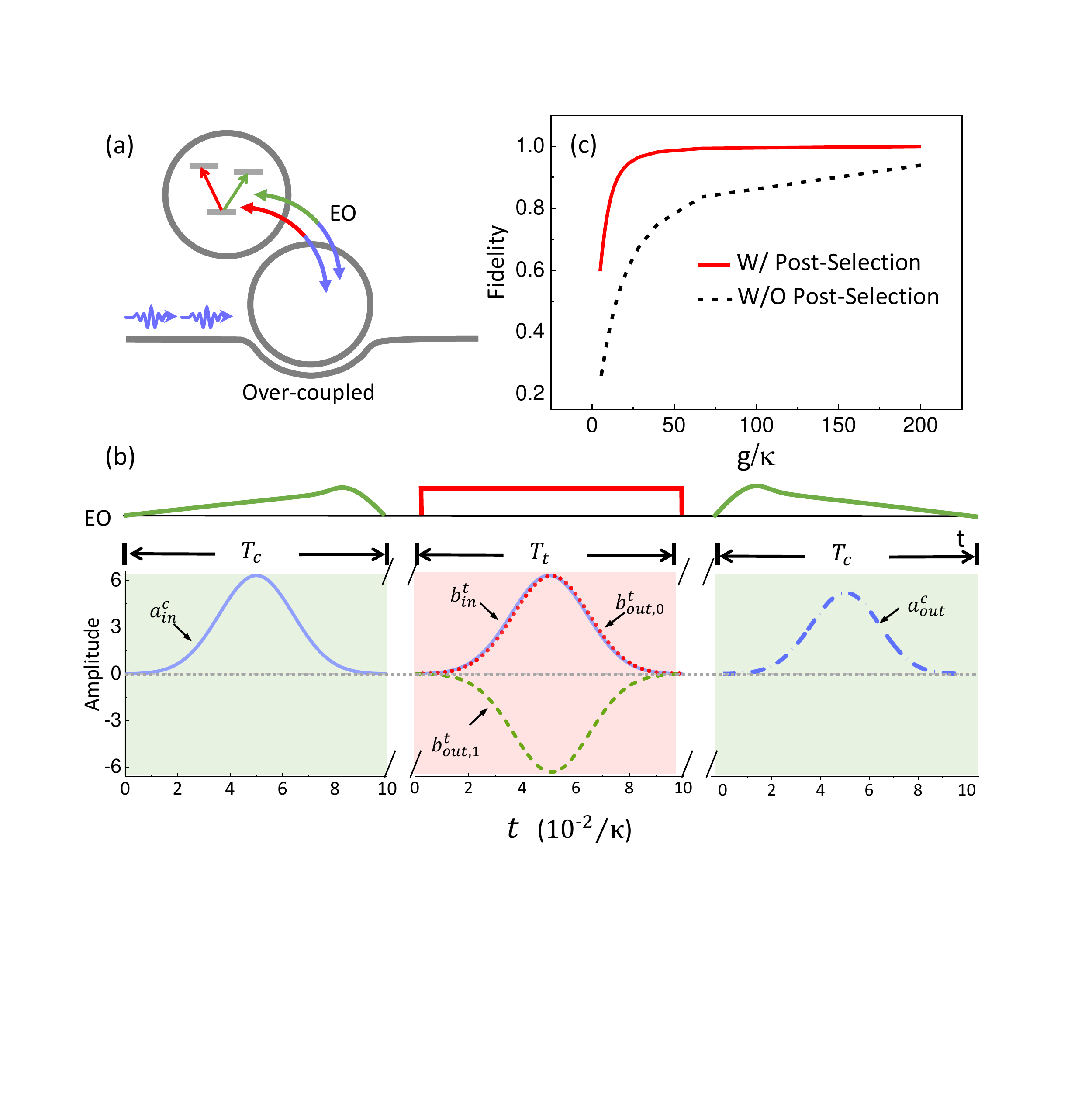}
\par\end{centering}
\caption{The photon-photon control-phase gate in a photonic molecule. (a) Schematic
of photonic molecule. The cavity coupled to the waveguide serves as
an active antenna shown in Fig.$\;$\ref{Fig2}(b). (b) Time sequence
of the EO drive for the quantum storage and retrieval of the photons.
At first, the control photon is stored into the artificial atom by
time-dependent driving $\Omega_{\mathrm{a}}(t)$. Then the target
photon is driven by another EO field with strength $\kappa_{\mathrm{a},0}\ll\Omega_{\mathrm{a}}\ll g$.
The output state of the target photon acquires a $\pi$ phase (shown
by the negativity of $b_{\mathrm{out},1}^{\mathrm{t}}$). After the
target photon passed, the control photon is retrieved from the artificial
atom. In the calculation, $\kappa_{\mathrm{d},1}=10^{3}\:\kappa_{\mathrm{d},0}$,
$\kappa_{\mathrm{d},0}=2\kappa$, the intrinsic decay rate of the
artificial atom $\kappa_{\mathrm{a},0}=\kappa_{\mathrm{b},0}=\kappa=1$.
The pulse shape of the control photon is a Guassian function $a_{\mathrm{in}}^{\mathrm{c}}=\alpha\left(\exp\left(-25(t/T_{\mathrm{c}}-0.5)-\exp(-6.25)\right)\right)$,
and similar for the target photon $b_{\mathrm{in}}^{\mathrm{t}}$.
$\alpha$ is the normalization factor. The duration of the two photons
are $T_{\mathrm{c}}=T_{\mathrm{t}}=0.1/\kappa$. (c) The gate fidelity
versus the nonlinear coupling strength $g$. In our scheme, the pulse
shape is maintained with high fidelity. The reason that degrades the
gate fidelity is mainly the dissipation of the atomic state, which
decays during the transmission of the target photon. By post-selecting
the two-photon events, the gate fidelity can be above $99\:\%$ for
$g>100$.}

\label{Fig3}
\end{figure}

As shown in Fig.$\;$\ref{Fig3}(a), an architecture based on the
photonic molecule is proposed to perform the $CZ$ gate. The architecture
composed of two microresonators and a bus waveguide, with one microresonator
engineered for artificial atom, while the other served as an antenna
to simultaneously couple with the artificial atom and waveguide {[}Fig.$\,$\ref{Fig2}(b){]}.
Since the two microresonators are made with the same $\chi^{(2)}$
material, thus the frequency of the antenna could be modulated by
EO effect and thereby the coupling between the waveguide and artificial
atom could be controlled in real-time. Such a photonic molecule can
be experimentally realized in LN integrated microrings and has been
demonstrated to store coherent lasers$\;$\cite{Zhang2019}. The Hamiltonian
of the system reads
\begin{eqnarray}
H & = & H_{\mathrm{AA}}+\omega_{\mathrm{d}}d^{\dagger}d+\sum_{A\in\left\{ a,b\right\} }\left(\Omega_{A}\left(t\right)Ad^{\dagger}+h.c.\right),\label{eq:Hamiltonian}
\end{eqnarray}
where $d$ is the bosonic operator for the ancillary mode (frequency
$\omega_{\mathrm{d}}$) in the antenna cavity, $A\in\left\{ \mathrm{a},\mathrm{b}\right\} $
is the operator for the modes in the artificial atom. By carefully
designing the geometry and tuning of the microresonators, the two
cavities have sightly different free spectral ranges. The mode $d$
is detuned from both mode $a$ and $b$, and the coupling $\Omega_{a,b}\left(t\right)$
between $d$ and $a$ or $b$ could be switched by controlling the
EO driving. For a strongly over-coupled antenna cavity with external
coupling rate $\kappa_{\mathrm{d},1}\gg\kappa_{\mathrm{d},0}$ ($\kappa_{\mathrm{d},0}$
is the intrinsic loss rate of the antenna cavity), it can be adiabatically
eliminated and mediates the effective coupling between the waveguide
and artificial atom. Considering the input field $A_{\mathrm{in}}\left(t\right)$
for $A\in\left\{ a,b\right\} $, the dynamics of the mode $A$ of
artificial atom follows$\;$\cite{walls2007quantum} 
\begin{eqnarray}
\frac{d}{dt}A & = & -i\left[A,H_{AA}\right]-\left(\widetilde{\kappa}_{A,0}+\widetilde{\kappa}_{A,1}\right)A+\sqrt{2\widetilde{\kappa}_{A,1}}A_{\mathrm{in}}\left(t\right),\label{eq:dyna}
\end{eqnarray}
with $\widetilde{\kappa}_{A,0}\approx\kappa_{A,0}+\left|\Omega_{A}\left(t\right)\right|^{2}\kappa_{\mathrm{d},0}/\left(\kappa_{\mathrm{d},0}+\kappa_{\mathrm{d},1}\right)^{2}$
and $\widetilde{\kappa}_{A,1}\approx\left|\Omega_{A}\left(t\right)\right|^{2}\kappa_{\mathrm{d},1}/\left(\kappa_{\mathrm{d},0}+\kappa_{\mathrm{d},1}\right)^{2}$
denoting the time-dependent effective external coupling rate controlled
by the antenna. By optimizing the shape and frequency of the EO field
$\Omega_{A}\left(t\right)$, the antenna can couple photons of different
shapes and frequencies in the waveguide to different energy levels.
Here, we perform the $CZ$ between identical photons as an instance.
To avoid crosstalk during the operation, the non-degenerate $\chi^{(2)}$
interaction is used and the EO drives have different frequencies for
two photons to couple them with different energy levels.

Figure$\;$\ref{Fig3}(b) shows the sequence of the $CZ$-gate scheme:
Initially, quantum states are encoded in control $\left(\left|\psi_{\mathrm{c}}\right\rangle =\alpha_{\mathrm{c}}\left|0\right\rangle +\beta_{\mathrm{c}}\left|1\right\rangle \right)$
and target $\left(\left|\psi_{\mathrm{t}}\right\rangle =\alpha_{\mathrm{t}}\left|0\right\rangle +\beta_{\mathrm{t}}\left|1\right\rangle \right)$
photons, which are temporally separated and send to the photonic molecule.
These two photons can only be coupled to the artificial atom if they
are prepared in state $|1\rangle$. Under appropriate EO drive, the
control photon of shape $a_{\mathrm{in}}^{\mathrm{c}}\left(t\right)$
is stored into the artificial atom $\left(\left|\psi_{\mathrm{AA}}\right\rangle =\alpha_{\mathrm{c}}\left|000\right\rangle +\beta_{\mathrm{c}}\left|100\right\rangle \right)$.
Subsequently, the target photon of shape $b_{\mathrm{in}}^{\mathrm{t}}\left(t\right)$
comes and another strong EO drive is applied. Due to the strong $\chi^{(2)}$
interaction, the transition $\left|100\right\rangle \rightarrow\left|110\right\rangle $
is blocked while $\left|000\right\rangle \rightarrow\left|010\right\rangle $
is allowed. Thus, the target photon will be reflected back to the
waveguide and acquires a $\pi$ phase depending on the state of the
artificial atom. The state becomes $\alpha_{\mathrm{c}}\left|000\right\rangle \otimes\left|\psi_{\mathrm{t}}\right\rangle +\beta_{\mathrm{c}}\left|100\right\rangle \otimes Z\left|\psi_{\mathrm{t}}\right\rangle $,
manifesting the $CZ$-gate with $Z$ denotes the Pauli matrix. Finally,
another EO drive is applied to retrieve the control photon back to
the waveguide. During the whole process, the two photons never meet
each other and the two-photon spontaneous emission is avoided.

For a given input Gaussian pulse shape $A_{\mathrm{in}}\left(t\right)$
of input photons, the maximum storage and retrieval efficiency 
\begin{equation}
\eta_{s}=\frac{\kappa_{\mathrm{d},1}}{\kappa_{\mathrm{d},0}+\kappa_{\mathrm{d},1}}\label{eq:efficiency}
\end{equation}
could be achieved by a carefully tailored driving pulse shape. For
retrieval, the optimal drive is derived as 
\begin{eqnarray}
\Omega(t) & = & -i\sqrt{\frac{\kappa_{\mathrm{d}}}{2}}\frac{A_{\mathrm{in}}\left(t\right)}{\sqrt{\int_{t}^{\infty}dt'|A_{\mathrm{in}}\left(t'\right)|^{2}}}\label{eq:shape}
\end{eqnarray}
(see the SM for details). The optimal drive for the storage process
is the time-reverse of that for the retrieval process. It is shown
that the quantum storage and retrieval efficiencies are independent
of the pulse shape \cite{storage}, and efficiency higher than $99.9\%$
is expected for the strongly over-coupled antenna due to the very
large coupling rate $\kappa_{\mathrm{d},1}/\kappa_{\mathrm{d},0}>10^{3}$
between the waveguide and cavity.

For the operations on both the control and target photons, there are
at most two excitations, thus the quantum dynamics of the system {[}Eq.$\;$(\ref{eq:dyna}){]}
can be solved in a truncated Fock state space. In Fig.$\;$\ref{Fig3}(b),
we calculated the evolution of the pulse shape for both the control
and target photon. It is shown that the shapes of the output fields
$b_{\mathrm{out},0}^{\mathrm{t}}$, $b_{\mathrm{out},1}^{\mathrm{t}}$
(subscript $0,1$ denotes the output conditioned on control state
$\left|0\right\rangle $,$\left|1\right\rangle $) for the target
photon and $a_{\mathrm{out}}^{\mathrm{t}}$ for the control photon
are nearly the same with the input pulses, indicating the high fidelity
of the quantum gate. The $\pi$ phase shift is indicated by the negativity
of $b_{\mathrm{out},1}^{\mathrm{t}}$. The relationship between the
gate fidelity and $g/\kappa$ ratio (the atom decay rate $\kappa_{A,0}=\kappa$)
is plotted in Fig.$\;$\ref{Fig3}(c) (dashed line). In our case,
the main imperfection affecting the gate fidelity is the lifetime
of the atomic states, corresponding to the cavity decay rate $\kappa$.
By normalizing the output state via post-selection, the gate fidelity
can be as high as $99\%$ for $g/\kappa\sim100$ {[}Fig.$\;$\ref{Fig3}(c),
solid line){]}.

Recalling that the internal states of an artificial atom could be
manipulated through coherent driving {[}Fig.$\;$\ref{Fig1}(c){]},
the artificial atom can also be used for mediating the entanglement
between photons. For example, the Duan-Kimble protocol$\;$\cite{DuanKimble}
and the Lindner-Rudolph protocol$\;$\cite{Lindner2009} could be
realized without photon storage and retrieval, but requiring the manipulation
of the internal state of the artificial atom. These protocols are
feasible in light of the fact that two-photon spontaneous emission
is avoided. Additionally, the higher excitation eigenstates of the
artificial atom can also be engineered to construct quantum gates
for the multi-level encoding of quantum states.

\emph{Discussion.- }From the above studies, the performances of artificial
atoms depend on the $g/\kappa$, as a benchmark for the cooperation
of nonlinear optical processes at the single-photon-level. Among various
available photonic platforms, the LN is appealing. It has a significant
$\chi^{(2)}$ susceptibility of $3\times10^{-11}\:\mathrm{m/V}$ and
excellent electro-optic property. Recent advances on its etching have
promoted the quality factor of integrated LN microcavity with a diameter
of tens of micrometers up to $10^{7}$$\;$\cite{zhang2017monolithic}.
Therefore, a coupling strength of $\sim5\:\mathrm{MHz}$ and a cavity
decay rate of $\sim20\:\mathrm{MHz}$ are feasible$~$\cite{LN2019},
which gives $g/\kappa$ be about $0.25$ and promises to observe the
quantum mechanical effect. More excitingly, crystalline LN offers
ultralow loss in the telecommunication band and the ultimate quality
factor of LN microcavityis higher than $10^{9}$~\cite{Zhang2019}.
By engineering the photonic mode volume and overlap, the value of
$g/\kappa$ could reach $100$. Furthermore, the excellent EO property
of LN enables dynamical and flexible control of the artificial atoms
and photonic structures.

There are also many other potential candidates for artificial atoms.
For example, ultra-high-efficiency SHG has been achieved in GaAs microresonators.
New materials with excellent nonlinear optics properties, such as
organic single crystals$\;$\cite{DAST} and two-dimensional materials$\;$\cite{Majumdar2015},
are recently developed and are potentially compatible with current
PIC platforms$\;$\cite{Li2018}. At the same time, new techniques
to engineer the $\chi^{(2)}$ nonlinearity have been developed . For
example, effective $\chi^{(2)}$ effect could be induced in centrosymmetric
materials, such as silicon waveguides by applying external bias electric
field$\;$\cite{Timurdogan2017}. Additionally, a scheme to synthesize
and amplify the nonlinear coupling strength with cascaded nonlinear
optical processes was proposed recently, and an effective $\chi^{(2)}$
interaction with $g/\kappa>1$ was predicted with optimal parameters
of LN microcavity$\;$\cite{amplification}.

\emph{Conclusion.- }We introduce an artificial atom on the photonic
integrated circuit by harnessing the cavity-enhanced optical $\chi^{(2)}$
nonlinearity in a microresonator. Such artificial atom preserves the
advantages of both atom and photonic cavity, and offers single-photon
level nonlinearity for deterministic quantum gates as well as experimental
compatibility and flexibility for scalable quantum devices on a chip.
Moreover, we propose a scheme based on the artificial atom to realize
high-fidelity two-photonic-qubit quantum control-phase gate by switching
the coupling channels between the artificial atom and continuum, which
also addresses the concerns raised by Shapiro$~$\cite{PhysRevA.73.062305,PhysRevA.90.062314}.
Due to recent advances in the development of low-loss thin-film LN
on insulator platform, the artificial atom with $g/\kappa\sim1$ is
currently feasible, and thereby quantum effects (such as photon blockade)
could be envisioned in a photonic chip maded with pure dielectrics.
The universal quantum gate set for quantum information processing
would also be achievable with further development of the fabrication
and materials. Our work opens an avenue to investigate the quantum
nonlinear photonics and is conducive to room-temperature, single-emitter-free
quantum information processing.

\emph{Note}: When finalizing this manuscript, there is a related theoretical
work posted on arXiv$\;$\cite{Heuck2019}.

\smallskip{}

\noindent \textbf{Acknowledgments}\\This work was funded by the National
Key R \& D Program (Grants No. 2016YFA0301300) and the National Natural
Science Foundation of China (Grant No.11874342, 11934012, 11922411,
11904316, and 11704370), and Anhui Initiative in Quantum Information
Technologies (AHY130200).

\bibliographystyle{aps}
\bibliography{chi2_gate}

\cleardoublepage{}

\onecolumngrid 
\global\long\def\thefigure{S\arabic{figure}}%
 \setcounter{figure}{0} 
\global\long\def\thepage{S\arabic{page}}%
 \setcounter{page}{1} 
\global\long\def\theequation{S.\arabic{equation}}%
 \setcounter{equation}{0} 
\setcounter{section}{0}
\begin{center}
\textbf{\Large{}SUPPLEMENTARY MATERIAL}{\Large\par}
\par\end{center}

\section{Anharmonic energy level structure of artificial atoms}

The cavity, which is designed for phase-matching condition that enables
non-degenerate $\chi^{(2)}$ interaction $\omega_{a}+\omega_{b}\rightarrow\omega_{c}$,
has a Hamiltonian of
\begin{eqnarray}
H & = & \omega_{\mathrm{a}}a^{\dagger}a+\omega_{\mathrm{b}}b^{\dagger}b+\omega_{\mathrm{c}}c^{\dagger}c+g_{\mathrm{nd}}\left(a^{\dagger}b^{\dagger}c+abc^{\dagger}\right),
\end{eqnarray}
where $a,b,c$ represents the bosonic operator for modes, $g_{\mathrm{nd}}$
is the nonlinear coupling strength. In the strong coupling regime
$g_{\mathrm{nd}}\gtrsim\kappa_{\mathrm{a}},\kappa_{\mathrm{b}},\kappa_{\mathrm{c}}$,
the system shows strong anhamonicity. For example, the state $|1_{\mathrm{a}}1_{\mathrm{b}}0_{\mathrm{c}}\rangle$
will couple to $|0_{\mathrm{a}}0_{\mathrm{b}}1_{\mathrm{c}}\rangle$
with a strength of $g_{\mathrm{nd}}$, and these two states will be
hybridized with each other, leading to a splitting of $2g_{\mathrm{nd}}$.
In this nonlinear cavity system, the eigenstates and eigenvalues in
each $N$-photon subspace could be solved independently. The energy
conversation requires
\begin{eqnarray}
N & = & N_{a}+N_{b}+2N_{c},
\end{eqnarray}
where $N_{i}$ is the number of excitation in mode $i$ ($i\in\left\{ \mathrm{a},\mathrm{b},\mathrm{c}\right\} $).
Therefore, the basis $\left\{ |\Psi_{N,m,n}\rangle=|m-1,n-1,N-m-n+2\rangle\right\} _{N}$
spans the $N$-photon subspace $H^{\left(N\right)}$, where $m,n\leq N$
are integers with $m+n-2<N$. The Hamiltonian is written in the matrix
form as
\begin{eqnarray}
H^{\left(N\right)} & = & \sum H_{m,n;m',n'}^{\left(N\right)}|\Psi_{N,m,n}\rangle\langle\Psi_{N,m',n'}|,
\end{eqnarray}
with the matrix element
\begin{eqnarray}
H_{m,n;m',n'}^{\left(N\right)} & = & \langle\Psi_{N,m',n'}|H|\Psi_{N,m,n}\rangle.
\end{eqnarray}
The eigenstates and eigenenergies are solved by diagonalizing the
matrix. Specially, we only consider the interaction between two photons,
thus truncate the dimension of the Fock state space to $2$ is enough.
The Hamiltonian in $2$-photon excitation is spanned by the eigenstate
$|\tilde{j}\rangle\in\left\{ \begin{array}{cc}
|1,1,0\rangle, & |0,0,1\rangle\end{array}\right\} $. For perfect phase matching, an appropriate rotating frame can be
chosen to transform $H$ to $H=g\left(a^{\dagger}b^{\dagger}c+abc^{\dagger}\right),$
\begin{eqnarray}
H_{ij}^{2} & = & \langle\tilde{j}|H|\tilde{i}\rangle_{2}\\
 & = & g_{\mathrm{nd}}\left(\delta(i-1,j)+\delta(i,j-1)\right),
\end{eqnarray}
where $\delta(x,y)$ is the Kronecker delta function. The energy levels
are obtained by diagolizing the matrix of the Hamiltonian. Therefore,
the degeneracy of the two-excitation states is removed by the nonlinear
interaction with energy levels split by $2g_{nd}$. If the cavity
is initially prepared in the state $|1_{a}0_{b}0_{c}\rangle$, the
photon resonant with mode $b$ can no longer efficiently induce the
transition from $|1_{a}0_{b}0_{c}\rangle$ to $|0_{a}0_{b}1_{c}\rangle$,
which is known as the photon-blockade effect. The energy levels of
second-harmonic generation $g_{\mathrm{d}}\left(a^{\dagger2}c+a^{2}c^{\dagger}\right)$
can be obtained following a similar calculation.

\section{Two-photon scattering by the artificial atom}

In this section, we investigate the two-photon transport through a
waveguide side-coupled to a cavity supporting $\chi^{(2)}$ interaction.
Figure 2(a) in the main text schematically illustrates a doubly-resonant
microcavity coupled with a bus waveguide. The two resonant modes in
the cavity are the fundamental (FM) mode $a$ and the second-harmonic
(SH) mode $c$ with frequency $\omega_{i}$ ($i\in\{a,c\}$). The
waveguide supports continuum states both in the fundamental and SH
frequency bands. Inside the cavity, modes $a$ and $c$ couple with
each other via the process of SH generation. The Hamiltonian of whole
system includes, the continuum states in the waveguide
\begin{eqnarray}
H_{W} & = & \int dxf^{\dagger}(x)\frac{1}{i}\frac{d}{dx}f(x)+\int dxs^{\dagger}(x)\frac{1}{i}\frac{d}{dx}s(x),
\end{eqnarray}
the localized states in the cavity
\begin{eqnarray}
H_{C} & = & \left(\omega_{a}-i\kappa_{a,0}\right)a^{\dagger}a+\left(\omega_{c}-i\kappa_{c,0}\right)c^{\dagger}c,
\end{eqnarray}
the nonlinear coupling inside the cavity
\begin{eqnarray}
H_{nl} & = & g_{\mathrm{d}}\left(a^{\dagger2}c+a^{2}c^{\dagger}\right),
\end{eqnarray}
the linear coupling between the waveguide and cavity
\begin{eqnarray}
H_{l} & = & V_{a}\int dx\delta(x)[f^{\dagger}(x)a+f(x)a^{\dagger}]+V_{c}\int dx\delta(x)[s^{\dagger}(x)c+s(x)c^{\dagger}],
\end{eqnarray}
where $x$ is coordinate along the waveguide, $f(x)$ ($s(x)$) and
$f^{\dagger}(x)$ ($s^{\dagger}(x)$) are the annihilation and creation
operators at position $x$ in the waveguide, $g_{\mathrm{d}}$ is
the nonlinear coupling strength of SH generation, $V_{a}$ ($V_{c}$)
is the coupling strength between waveguide and cavity in the fundamental
(SH) frequency band, $\kappa_{a,0}$ ($\kappa_{c,0}$) is the intrinsic
decay rates of the cavity mode. Since the coupling near the narrow
resonant frequency window is considered, the dispersions of the coupling
strengths $g_{\mathrm{d}}$, $V_{a}$ and $V_{c}$ are neglected.
The input state is initially prepared in a two-photon state at the
fundamental frequencies. A general form of two-excitation state can
be written as
\begin{eqnarray}
|E_{2}\rangle & = & \int dx_{1}dx_{2}g(x_{1},x_{2})f^{\dagger}(x_{1})f^{\dagger}(x_{2})|0\rangle+\int dxm(x)f^{\dagger}(x)|0\rangle+\nonumber \\
 &  & \int dxh(x)s^{\dagger}(x)|0\rangle+f_{a}\frac{1}{\sqrt{2}}a^{\dagger}a^{\dagger}|0\rangle+f_{c}c^{\dagger}|0\rangle.
\end{eqnarray}
Following the Schrodinger equation $H|E_{2}\rangle=E_{2}|E_{2}\rangle$
($E_{2}=k_{1}+k_{2}$, $k_{i}$ is the frequency of the $i$-th photon),
one gets
\begin{eqnarray}
\left(-i\frac{\partial}{\partial x_{1}}-i\frac{\partial}{\partial x_{2}}-E_{2}\right)g(x_{1},x_{2})+\frac{V_{a}}{2}[\delta(x_{1})m(x_{2})+\delta(x_{2})m(x_{1})] & = & 0,\label{eq:two-ea-1}\\
\left(-i\frac{d}{dx}+\omega_{a}-i\kappa_{a,0}-E_{2}\right)m(x)+V_{a}[g(x,0)+g(0,x)]+\sqrt{2}V_{a}f_{a}\delta(x) & = & 0,\\
\left(-i\frac{d}{dx}-E_{2}\right)h(x)+V_{c}f_{c}\delta(x) & = & 0,\label{eq:two-ea}\\
\left(2\omega_{a}-2i\kappa_{a,0}-E_{2}\right)f_{a}+\sqrt{2}g_{\mathrm{d}}f_{c}+\sqrt{2}V_{a}m(0) & = & 0,\\
\left(\omega_{c}-i\kappa_{c,0}-E_{2}\right)f_{c}+\sqrt{2}g_{\mathrm{d}}f_{a}+V_{c}h(0) & = & 0.\label{eq:two-ea-5}
\end{eqnarray}
The discontinuous functions fulfills
\begin{eqnarray}
g(x,0) & = & g(0,x),\\
g(x,0) & = & [g(x,0^{+})+g(x,0^{-})]/2,\\
m(0) & = & [m(0^{+})+m(0^{-})]/2,\\
h(0) & = & [h(0^{+})+h(0^{-})]/2,\\
\frac{d}{dx}m(x)|_{x=0} & = & [m(0^{+})-m(0^{-})]\delta(x).
\end{eqnarray}
Then Eqs.(\ref{eq:two-ea-1})-(\ref{eq:two-ea-5}) are equivalent
to
\begin{eqnarray}
\left(-i\frac{\partial}{\partial x_{1}}-i\frac{\partial}{\partial x_{2}}-E_{2}\right)g(x_{1},x_{2}) & = & 0,\\
\frac{2i}{V_{a}}[g(0^{+},x)-g(0^{-},x)] & = & m(x),\label{eq:2}\\
\left(-i\frac{d}{dx}+\omega_{a}-i\kappa_{a,0}-E_{2}\right)m(x)+V_{a}[g(x,0^{+})+g(0^{-},x)] & = & 0,\label{eq:3}\\
\frac{i}{\sqrt{2}V_{a}}[m(0^{+})-m(0^{-})] & = & f_{a},\\
\left(-i\frac{d}{dx}-E_{2}\right)h(x) & = & 0,\\
\frac{i}{V_{c}}[h(0^{+})-h(0^{-})] & = & f_{c},\\
\left(2\omega_{a}-2i\kappa_{a,0}-E_{2}\right)f_{a}+\sqrt{2}g_{\mathrm{d}}f_{c}+\sqrt{2}V_{a}m(0) & = & 0,\\
\left(\omega_{c}-i\kappa_{c,0}-E_{2}\right)f_{c}+\sqrt{2}g_{\mathrm{d}}f_{a}+V_{c}h(0) & = & 0.
\end{eqnarray}
Combine Eqs.$\,$(\ref{eq:2}) and (\ref{eq:3}), we get
\begin{eqnarray}
\left(-i\frac{d}{dx}+\omega_{a}-i\kappa_{a,0}-i\kappa_{a,1}-E_{2}\right)g(0^{+},x) & = & \left(-i\frac{d}{dx}+\omega_{a}-i\kappa_{a,0}+i\kappa_{a,1}-E_{2}\right)g(0^{-},x).\label{eq:region}
\end{eqnarray}
Adopting the method in Ref.$\,$\cite{Zheng2013}, $g(x_{1},x_{2})$
is divided to three regions: $g_{1}(x_{1},x_{2})$ for $x_{1}\leq x_{2}<0$,
$g_{2}(x_{1},x_{2})$ for $x_{1}<0<x_{2}$ and $g_{3}(x_{1},x_{2})$
for $0<x_{1}\leq x_{2}$. For two-photon monochromatic wave input
state
\begin{eqnarray}
g_{1}(x_{1},x_{2}) & = & \frac{1}{2\text{\ensuremath{\sqrt{2}\pi}}}\left(e^{ik_{1}x_{1}}e^{ik_{2}x_{2}}+e^{ik_{2}x_{1}}e^{ik_{1}x_{2}}\right),
\end{eqnarray}
$g(x_{1},x_{2})$ in the other two regions are derived from Eq.$\,$(\ref{eq:region})
as
\begin{eqnarray}
g_{2}(x_{1},x_{2}) & = & \frac{1}{2\text{\ensuremath{\sqrt{2}\pi}}}\left(t_{k_{2}}e^{i(k_{1}x_{1}+k_{2}x_{2})}+t_{k_{1}}e^{i(k_{2}x_{1}+k_{1}x_{2})}\right),\\
g_{3}(x_{1},x_{2}) & = & \frac{1}{2\text{\ensuremath{\sqrt{2}\pi}}}t_{k_{1}}t_{k_{2}}\left(e^{i(k_{1}x_{1}+k_{2}x_{2})}+e^{i(k_{2}x_{1}+k_{1}x_{2})}\right)+Be^{-i(\omega_{a}-i\kappa_{a,0}-i\kappa_{a,1})(x_{2}-x_{1})}e^{i(k_{1}+k_{2})x_{2}},
\end{eqnarray}
with
\begin{eqnarray}
t_{k} & = & \frac{k-\omega_{a}+i\kappa_{a,0}-i\kappa_{a,1}}{k-\omega_{a}+i\kappa_{a,0}+i\kappa_{a,1}}.
\end{eqnarray}
Eliminating $f_{a}$ and $f_{c}$ ,
\begin{eqnarray}
\frac{2i}{V_{a}}[\alpha_{a}^{-}g(0^{+},0^{+})-(\alpha_{a}^{-}+\alpha_{a}^{+})g(0^{-},0^{+})+\alpha_{a}^{+}g(0^{-},0^{-})]+2g_{\mathrm{d}}\frac{V_{a}}{V_{c}}[h(0^{+})-h(0^{-})] & = & 0,\\
\alpha_{b}^{-}h(0^{+})-\alpha_{b}^{+}h(0^{-})+\frac{2ig_{\mathrm{d}}V_{c}}{V_{a}^{2}}[g(0^{+},0^{+})-2g(0^{-},0^{+})+g(0^{-},0^{-})] & = & 0,
\end{eqnarray}
with $\alpha_{a}^{\pm}=2\left(\omega_{a}-i\kappa_{a,0}\pm i\kappa_{a,1}\right)-k_{1}-k_{2}$,
$\alpha_{c}^{\pm}=\left(\omega_{c}-i\kappa_{c,0}\pm i\kappa_{c,1}\right)-k_{1}-k_{2}$,
$\kappa_{i,1}=V_{i}^{2}/2$. Using the values at the discontinuous
points, 
\begin{eqnarray}
g(0^{+},0^{+})=g_{3}(0,0) & = & \frac{1}{\sqrt{2}\pi}t_{k_{1}}t_{k_{2}}+B,\\
g(0^{-},0^{+})=g_{2}(0,0) & = & \frac{1}{2\sqrt{2}\pi}(t_{k_{1}}+t_{k_{2}}),\\
g(0^{-},0^{-})=g_{1}(0,0) & = & \frac{1}{\sqrt{2}\pi},\\
h(0^{+}) & = & \frac{1}{\sqrt{2\pi}}C,\\
h(0^{-}) & = & 0,
\end{eqnarray}
one finally obtains
\begin{eqnarray}
C & = & \frac{ig_{\mathrm{d}}\sqrt{2\kappa_{c,1}}\left(\alpha_{a}^{+}-\alpha_{a}^{-}\right)\left(2-t_{k_{1}}-t_{k_{2}}\right)}{2\sqrt{\pi}\kappa_{a,1}\left(\alpha_{a}^{-}\alpha_{c}^{-}-2g_{d}^{2}\right)}\nonumber \\
 & = & -\frac{2g_{\mathrm{d}}\sqrt{2\kappa_{c,1}}\left(2-t_{k_{1}}-t_{k_{2}}\right)}{\sqrt{\pi}\left(\alpha_{a}^{-}\alpha_{c}^{-}-2g_{\mathrm{d}}^{2}\right)},\\
B & = & \frac{4g_{\mathrm{d}}^{2}\left(t_{k_{1}}-1\right)\left(t_{k_{2}}-1\right)+\alpha_{a}^{+}\alpha_{c}^{-}\left(t_{k_{1}}+t_{k_{2}}-2\right)+\alpha_{a}^{-}\alpha_{c}^{-}\left(t_{k_{1}}+t_{k_{2}}-2t_{k_{1}}t_{k_{2}}\right)}{2\sqrt{2}\pi\left(\alpha_{a}^{-}\alpha_{c}^{-}-2g_{\mathrm{d}}^{2}\right)}\nonumber \\
 & = & \frac{\sqrt{2}g_{\mathrm{d}}^{2}\left(1-t_{k_{1}}\right)\left(1-t_{k_{2}}\right)}{\pi\left(\alpha_{a}^{-}\alpha_{c}^{-}-2g_{\mathrm{d}}^{2}\right)}.
\end{eqnarray}
The expressions can be simplified as
\begin{eqnarray}
C & = & -\frac{2g_{\mathrm{d}}\sqrt{2\kappa_{c,1}}}{\sqrt{\pi}\left[2\omega_{a}-k_{1}-k_{2}-2i\kappa_{a,tot}\right]\left[\omega_{c}-k_{1}-k_{2}-i\kappa_{c,tot}\right]-2g_{d}^{2}}\left(\frac{2i\kappa_{a,1}}{k_{1}-\omega_{a}+i\kappa_{a,tot}}+\frac{2i\kappa_{a,1}}{k_{2}-\omega_{a}+i\kappa_{a,tot}}\right),\\
B & = & \frac{\sqrt{2}g_{d}^{2}}{\pi\left[2\omega_{a}-k_{1}-k_{2}-2i\kappa_{a,tot}\right]\left[\omega_{c}-k_{1}-k_{2}-i\kappa_{c,tot}\right]-2g_{\mathrm{d}}^{2}}\frac{2i\kappa_{a,1}}{k_{1}-\omega_{a}+i\kappa_{a,tot}}\frac{2i\kappa_{a,1}}{k_{2}-\omega_{a}+i\kappa_{a,tot}}.
\end{eqnarray}
where $\kappa_{i,tot}=\kappa_{i,0}+\kappa_{i,1}$. The wave-functions
of two-photon input and output states are computed by the Lippmann-Schwinger
equations$\:$\cite{lipp}, 
\begin{eqnarray}
\phi_{in}(x_{1},x_{2}) & = & \langle x_{1},x_{2}|E_{2}\rangle-\langle x_{1},x_{2}|\frac{1}{k_{1}+k_{2}-H_{0}+i0^{+}}H_{int}|E_{2}\rangle\nonumber \\
 & = & g_{1}(x_{1},x_{2})\theta(x_{2}-x_{1})+g_{1}(x_{2},x_{1})\theta(x_{1}-x_{2}),\\
\phi_{out}(x_{1},x_{2}) & = & \langle x_{1},x_{2}|E_{2}\rangle-\langle x_{1},x_{2}|\frac{1}{k_{1}+k_{2}-H_{0}-i0^{+}}H_{int}|E_{2}\rangle\nonumber \\
 & = & g_{3}(x_{1},x_{2})\theta(x_{2}-x_{1})+g_{3}(x_{2},x_{1})\theta(x_{1}-x_{2}).
\end{eqnarray}
Then the two-photon scattering matrix is obtained by calculating the
wave-function overlap between the output state and the two photon
plain wave state,
\begin{eqnarray}
S(p_{1},p_{2};k_{1},k_{2}) & = & \int dx_{1}dx_{2}\frac{1}{2\sqrt{2}\pi}[e^{-i(p_{1}x_{1}+p_{2}x_{2})}+e^{-i(p_{2}x_{1}+p_{1}x_{2})}]\phi_{out}(x_{1},x_{2}).
\end{eqnarray}
The transmission plane wave term is
\begin{eqnarray}
S_{plane} & = & \frac{1}{8\pi^{2}}\int dx_{1}dx_{2}[e^{-i(p_{1}-k_{1})x_{1}}e^{-i(p_{2}-k_{2})x_{2}}+e^{-i(p_{1}-k_{2})x_{1}}e^{-i(p_{2}-k_{1})x_{2}}\nonumber \\
 &  & +e^{-i(p_{1}-k_{2})x_{2}}e^{-i(p_{2}-k_{1})x_{1}}+e^{-i(p_{2}-k_{2})x_{1}}e^{-i(p_{1}-k_{1})x_{2}}]\nonumber \\
 & = & t_{k_{1}}t_{k_{2}}[\delta(p_{1}-k_{1})\delta(p_{2}-k_{2})+\delta(p_{1}-k_{2})\delta(p_{2}-k_{1})]\nonumber \\
 & = & S_{p_{1}k_{1}}S_{p_{2}k_{2}}+S_{p_{1}k_{2}}S_{p_{2}k_{1}}.
\end{eqnarray}
The bound state term is
\begin{eqnarray}
S_{bound} & = & \frac{B}{2\sqrt{2}\pi}\int dx_{1}dx_{2}[e^{-i(p_{1}x_{1}+p_{2}x_{2})}+e^{-i(p_{2}x_{1}+p_{1}x_{2})}]e^{-i(\omega_{a}-i\kappa_{a,0}-i\kappa_{a,1})(x_{2}-x_{1})}e^{i(k_{1}+k_{2})x_{2}}\theta(x_{2}-x_{1})+\nonumber \\
 &  & \frac{B}{2\sqrt{2}\pi}\int dx_{1}dx_{2}[e^{-i(p_{1}x_{1}+p_{2}x_{2})}+e^{-i(p_{2}x_{1}+p_{1}x_{2})}]e^{-i(\omega_{a}-i\kappa_{a,0}-i\kappa_{a,1})(x_{1}-x_{2})}e^{i(k_{1}+k_{2})x_{1}}\theta(x_{1}-x_{2}).
\end{eqnarray}
The integral of the first term is simplified by replacing $x_{2}$
by $x_{3}+x_{1}$ and similar for the second term. We get
\begin{eqnarray}
S_{bound,1} & = & \frac{B}{2\sqrt{2}\pi}\int dx_{1}dx_{3}e^{-(\kappa_{a,0}+\kappa_{a,1})x_{3}}[e^{-i(p_{1}+p_{2}-k_{1}-k_{2})x_{1}}e^{-i(p_{2}+\omega_{a}-k_{1}-k_{2})x_{3}}\nonumber \\
 &  & +e^{-i(p_{1}+p_{2}-k_{1}-k_{2})x_{1}}e^{-i(p_{1}+\omega_{a}-k_{1}-k_{2})x_{3}}]\theta(x_{3})\nonumber \\
 & = & \frac{B}{2\sqrt{2}\pi}\times2\pi[\frac{1}{(\kappa_{a,0}+\kappa_{a,1})+i(p_{2}+\omega_{a}-k_{1}-k_{2})}\delta(p_{1}+p_{2}-k_{1}-k_{2})\nonumber \\
 &  & +\frac{1}{(\kappa_{a,0}+\kappa_{a,1})+i(p_{1}+\omega_{a}-k_{1}-k_{2})}\delta(p_{1}+p_{2}-k_{1}-k_{2})]\nonumber \\
 & = & \frac{B}{\sqrt{2}}[\frac{i}{p_{1}-\omega_{a}+i(\kappa_{a,0}+\kappa_{a,1})}+\frac{i}{p_{2}-\omega_{a}+i(\kappa_{a,0}+\kappa_{a,1})}]\delta(p_{1}+p_{2}-k_{1}-k_{2}).
\end{eqnarray}
Due to the symmetry of the wavefunction
\begin{eqnarray}
S_{bound} & = & 2S_{bound,1}\nonumber \\
 & = & \sqrt{2}B[\frac{i}{p_{1}-\omega_{a}+i(\kappa_{a,0}+\kappa_{a,1})}+\frac{i}{p_{2}-\omega_{a}+i(\kappa_{a,0}+\kappa_{a,1})}]\delta(p_{1}+p_{2}-k_{1}-k_{2})\nonumber \\
 & = & \sqrt{2}B[\frac{i}{p_{1}-\omega_{a}+i(\kappa_{a,0}+\kappa_{a,1})}+\frac{i}{p_{2}-\omega_{a}+i(\kappa_{a,0}+\kappa_{a,1})}]\delta(p_{1}+p_{2}-k_{1}-k_{2})\nonumber \\
 & = & \frac{B}{\sqrt{2}\kappa_{a,1}}(2-t_{p_{1}}-t_{p_{2}})\delta(p_{1}+p_{2}-k_{1}-k_{2}).
\end{eqnarray}
The total scattering matrix is 
\begin{eqnarray}
S(p_{1},p_{2};k_{1},k_{2}) & = & t_{k_{1}}t_{k_{2}}[\delta(p_{1}-k_{1})\delta(p_{2}-k_{2})+\delta(p_{1}-k_{2})\delta(p_{2}-k_{1})]\\
 &  & +\frac{B}{\sqrt{2}\kappa_{a,1}}(2-t_{p_{1}}-t_{p_{2}})\delta(p_{1}+p_{2}-k_{1}-k_{2}).
\end{eqnarray}
According to the scattering matrix, one can compute the output spectrum
of the two-photon state as
\begin{eqnarray}
\alpha_{out}(p_{1},p_{2}) & = & \int\int\alpha_{in}(k_{1},k_{2})S(p_{1},p_{2};k_{1},k_{2})dk_{1}dk_{2},
\end{eqnarray}
where $\alpha_{in}(k_{1},k_{2})$ is the input spectrum.

From the frequency conversion point of view, one might care about
the conversion efficiency of the two-photon second-harmonic generation.
The two-photon to one-photon conversion matrix is
\begin{eqnarray}
S(p;k_{1},k_{2}) & = & \int dxe^{-ipx}\frac{1}{\sqrt{2\pi}}C(k_{1}+k_{2})e^{i(k_{1}+k_{x})x}\nonumber \\
 & = & \frac{1}{\sqrt{2\pi}}C(k_{1}+k_{2})\int dxe^{-i(p-k_{1}-k_{2})x}\nonumber \\
 & = & C(k_{1}+k_{2})\delta(p-k_{1}-k_{2}).
\end{eqnarray}
The output spectrum of the second-harmonic photon is 
\begin{eqnarray}
\beta_{out}(p) & = & \int\int\alpha_{in}(k_{1},k_{2})S(p;k_{1},k_{2})dk_{1}dk_{2}\nonumber \\
 &  & \int\int\alpha_{in}(k_{1},k_{2})C(k_{1}+k_{2})\delta(p-k_{1}-k_{2})dk_{1}dk_{2}.
\end{eqnarray}
Then, we get the conversion efficiency
\begin{eqnarray}
\eta & = & \frac{\int|\beta_{out}(p)|^{2}dp}{2\int\int|\alpha_{in}(k_{1},k_{2})|^{2}dk_{1}dk_{2}}.
\end{eqnarray}

\section{Control phase gate via photonic molecule}

\subsection{Coupled microresonator}

In our scheme, an active antenna is used to couple the traveling photons
and the artificial atom, as shown by Fig.$\:$3 in the main text.
For artificial atom supporting mode $a,b$ (with resonant frequency
$\omega_{\mathrm{a}},\omega_{\mathrm{b}}$), drive field with frequency
$\omega_{\mathrm{EO}}=\omega_{A}-\omega_{0}$ can be used to induce
the coupling between photons of frequency $\omega_{0}$ and the mode
$A$ ($A\in\left\{ \mathrm{a},\mathrm{b}\right\} $). Such an antenna
can be physically realized with a microcavity and electro-optic (EO)
effect \cite{Zhang2019}. A microwave driving is applied on the two
cavities which shifts the two cavity resonant frequencies to opposite
directions, with the photonic molecular Hamiltonian as
\begin{eqnarray}
H & = & \omega_{m,1}m_{1}^{\dagger}m_{1}+\omega_{m,2}m_{2}^{\dagger}m_{2}+J\left(m_{1}^{\dagger}m_{2}+m_{1}m_{2}^{\dagger}\right)+G\cos(\omega_{\mathrm{EO}}t)\left(m_{1}^{\dagger}m_{1}-m_{2}^{\dagger}m_{2}\right),
\end{eqnarray}
where $\omega_{m,i}$ is the resonant frequency of the cavity mode
$m_{i}$, $J$ is the linear coupling strength between the two cavities,
$G$ is the coupling strength due to $\chi^{(2)}$ interaction between
photonic mode and electric field, i.e. the EO drive. Denoting $\omega_{0}=\frac{\omega_{m,1}+\omega_{m,2}}{2}$,
$\delta\omega=\frac{\omega_{m,1}-\omega_{m,2}}{2}$, $\tan\theta=\frac{J}{\delta\omega}$
and making the linear transformation 
\begin{eqnarray}
b & = & \cos\frac{\theta}{2}m_{1}+\sin\frac{\theta}{2}m_{2},\\
d & = & -\sin\frac{\theta}{2}m_{1}+\cos\frac{\theta}{2}m_{2},
\end{eqnarray}
the Hamiltonian transforms to
\begin{eqnarray}
H & = & \omega_{b}b^{\dagger}b+\omega_{d}d^{\dagger}d+G\cos(\omega_{\mathrm{EO}}t)\sin\theta\left(b^{\dagger}d+bd^{\dagger}\right)+G\cos(\omega_{\mathrm{EO}}t)\cos\theta\left(b^{\dagger}b-d^{\dagger}d\right),
\end{eqnarray}
where $\omega_{b}=\omega_{0}+\sqrt{\delta\omega^{2}+J^{2}}$, $\omega_{d}=\omega_{0}-\sqrt{\delta\omega^{2}+J^{2}}$.
For modulation frequency $\omega_{m}$ much larger than the cavity
decay rates $\kappa_{b},\kappa_{d}$ of the supermodes $b,d$, the
sidebands of mode $b$ and $d$ generated by the frequency modulated
term $\Omega\cos(\omega_{\mathrm{EO}}t)\cos\theta\left(b^{\dagger}b-dd^{\dagger}\right)$
are no longer on resonance, which can be neglected safely. Therefore,
the Hamiltonian reduces to
\begin{eqnarray}
H & = & \omega_{b}b^{\dagger}b+\omega_{d}d^{\dagger}d+G\cos(\omega_{\mathrm{EO}}t)\sin\theta\left(b^{\dagger}d+bd^{\dagger}\right).
\end{eqnarray}
 If only mode $m_{2}$ couples with the waveguide with rate $\sqrt{2\kappa_{m,2}}$,
mode $b$ and $d$ experiences effective coupling strength of $\sqrt{2\kappa_{m,2}}\cos\frac{\theta}{2}$
and $\sqrt{2\kappa_{m,2}}\sin\frac{\theta}{2}$, respectively. For
the case of large detuning $\delta\omega\gg\left\{ J,\:\kappa\right\} $,
$b\approx m_{1}$, $d\approx m_{2}$, mode $b$ is approximately decoupled
from the waveguide and its lifetime is only limited by the intrinsic
loss. In the rotating frame of $\omega_{b}b^{\dagger}b+\omega_{d}d^{\dagger}d$
and apply the rotating wave approximation, the Hamiltonian is
\begin{eqnarray}
H & = & \frac{1}{2}\Omega\sin\theta\left(b^{\dagger}de^{i\delta}+bd^{\dagger}e^{-i\delta}\right),
\end{eqnarray}
where $\delta=\omega_{d}-(\omega_{b}-\omega_{d})$. Transforming the
Hamiltonian further to the frame of $\frac{\delta}{2}c^{\dagger}c-\frac{\delta}{2}d^{\dagger}d$,
we obtain the time-independent form
\begin{eqnarray}
H & = & -\frac{\delta}{2}b^{\dagger}b+\frac{\delta}{2}d^{\dagger}d+\frac{1}{2}G\sin\theta\left(b^{\dagger}d+bd^{\dagger}\right).
\end{eqnarray}
In this way, we obtain the coupling between mode $b$ ($m_{1}$) in
the artificial atom cavity and the mode $d$ ($m_{2}$) in the antenna
cavity. The frequency difference between mode $d$ and $b$ is compensated
by the frequency of the EO drive. For input photon with pulse shape
$A_{in}\left(t\right)$, it is rapidly coupled into the antenna cavity
due to the large waveguide-cavity coupling rate and then coupled to
the artificial atom by a time-dependent $G\left(t\right)$.For two
incident photons with the same frequency, we use different $\omega_{\mathrm{EO}}$
to couple them with energy levels of different frequencies without
crosstalk.

\subsection{Quantum storage of control photon pulse}

We use the photonic molecule introduced in the last section for the
storage and readout of the control photon from the waveguide to the
artificial atom. The total Hamiltonian of the system is 
\begin{eqnarray}
H & = & \sum_{j}\omega_{j}j^{\dagger}j+g_{\mathrm{nd}}abc^{\dagger}+\Omega\left(t\right)bd^{\dagger}+h.c.
\end{eqnarray}
where $j\in\{a,b\}$ represents the mode in the atomic cavity in the
fundamental frequency band, $c$ is the mode in the atomic cavity
in the SH frequency band, $d$ is the mode in the antenna cavity.
The dynamics of the cavity field operator $A$ follow 
\begin{eqnarray}
\frac{d}{dt}A & = & -i\left[A,H\right]-\kappa_{A}A+\sqrt{2\kappa_{A,1}}A_{in}\left(t\right).
\end{eqnarray}
The information of the pulse shape is contained in the time dependent
term $A_{in}\left(t\right)$.

Here, we investigate the quantum storage of single photon with the
photonic molecule. Without the control photon, the strong $\chi^{(2)}$
interaction is forbidden, the Hamiltonian for photon storage is
\begin{eqnarray}
H & = & \delta_{b}b^{\dagger}b+\delta_{d}d^{\dagger}d+\Omega b^{\dagger}d+\Omega^{*}bd^{\dagger}.\label{eq:storage}
\end{eqnarray}
For single-photon excitation, the state can be described by a pure
state with a general form of the superposition of Fock states. The
single-photon state is expressed as
\begin{eqnarray}
|\psi\rangle_{db} & = & c_{10}|10\rangle+c_{01}|01\rangle,
\end{eqnarray}
where $|mn\rangle=|m\rangle_{d}\otimes|n\rangle_{b}$. The dynamics
follows
\begin{eqnarray}
\frac{d}{dt}c_{mn} & = & \frac{d}{dt}\langle mn|\psi\rangle\nonumber \\
 & = & \langle mn|\frac{d}{dt}\psi\rangle\nonumber \\
 & = & -i\langle mn|H_{\mathrm{eff}}|\psi\rangle\nonumber \\
 & = & \left(-i\delta_{b}-\kappa_{b}\right)\sum_{m'n'}c_{m'n'}\langle mn|b^{\dagger}b|m'n'\rangle+\left(-i\delta_{d}-\kappa_{d}\right)\sum_{m'n'}c_{m'n'}\langle mn|d^{\dagger}d|m'n'\rangle\nonumber \\
 &  & -i\left(\Omega\sum_{m'n'}c_{m'n'}\langle mn|b^{\dagger}d|m'n'\rangle+\Omega^{*}\sum_{m'n'}c_{m'n'}\langle mn|bd^{\dagger}|m'n'\rangle\right)\nonumber \\
 & = & \alpha_{b}nc_{mn}+\alpha_{d}mc_{mn}-i\Omega\sqrt{(m+1)n}c_{m+1,n-1}-i\Omega^{*}\sqrt{m(n+1)}c_{m-1,n+1},
\end{eqnarray}
where $\alpha_{i}=-i\delta_{i}-\kappa_{i}$, $H_{\mathrm{eff}}$ is
the effective Hamiltonian that includes the mode dissipation, which
is obtained by replacing $\delta_{i}$ by $\alpha_{i}$ in Eq.$\:$\ref{eq:storage}.
Under single-photon pulse excitation of a perfectly phase-matched
cavity $\alpha_{i}=\kappa_{i}$, the dynamics of the intracavity and
output fields follow 
\begin{eqnarray}
\frac{d}{dt}c_{10} & = & -\kappa_{d}c_{10}-i\Omega^{*}\left(t\right)c_{01}+\sqrt{2\kappa_{d,1}}\phi_{in}(t),\label{eq:cme1-1}\\
\frac{d}{dt}c_{01} & = & -\kappa_{b}c_{01}-i\Omega\left(t\right)c_{10},\label{eq:cme1-2}\\
\phi_{out} & = & \phi_{in}-\sqrt{2\kappa_{c,1}}c_{10},\label{eq:cme1-3}
\end{eqnarray}
where $\phi_{in}(t)$ ($\phi_{out}(t)$) represents the pulse shape
of the input (output) single photon.

\subsubsection{Figure of merit}

We assume that both cavities are in the ground state, so $c_{10}=c_{01}=0$
at $t=0$. We also assume that the input field excitation $\phi_{in}(t)$
only has nonzero values in time interval $[0,T]$. Our goal is to
store the input field to $|01\rangle$ and retrieve the photon out
of the cavity after. During the storage process, $c_{10}(0)=c_{01}(0)=0$
and the input field fulfills $\int_{0}^{T}dt|\phi_{in}(t)|^{2}=1$.
The storage efficiency is given by
\begin{eqnarray}
\eta_{s} & = & \frac{\mathrm{Population}\:\mathrm{of\:|01\rangle}}{\mathrm{Number\:\mathrm{of\:\mathrm{incoming}\:\mathrm{photons}}}}\nonumber \\
 & = & |c_{01}(T)|^{2}.
\end{eqnarray}
For the retrieval process, we set $c_{10}(t_{r})=0$, $c_{01}(t_{r})=1$
with $t_{r}$ being the start time of the retrieve. The retrieval
efficiency is given by
\begin{eqnarray}
\eta_{r} & = & \frac{\mathrm{Number}\:\mathrm{of\:\mathrm{retrieved}\:\mathrm{photons}}}{\mathrm{Population}\:\mathrm{of\:|01\rangle}}\nonumber \\
 & = & \int_{t_{r}}^{\infty}|\phi_{out}(t)|^{2}dt.\label{eq:retrive}
\end{eqnarray}

\subsubsection{Photon Retrieval}

First, we investigate the retrieval process with $\phi_{in}(t)=0$
for $t>t_{r}$. In our case, we have assumed that the second cavity
has a quality factor much larger than the first one which couples
with a bus waveguide, thus it is reasonable to make a approximation
$\kappa_{b}\approx0$. From Eqs.~(\ref{eq:cme1-1})-(\ref{eq:cme1-2}),
\begin{eqnarray}
\frac{d}{dt}\left(|c_{10}|^{2}+|c_{01}|^{2}\right) & = & -2\kappa_{d}|c_{10}|^{2}.
\end{eqnarray}
Using Eq.~(\ref{eq:cme1-3}) and Eq.~(\ref{eq:retrive}), the retrieval
efficiency is derived as
\begin{eqnarray}
\eta_{s} & = & 2\kappa_{d,1}\int_{t_{r}}^{\infty}dt|c_{10}|^{2}\nonumber \\
 & = & -\frac{\kappa_{d,1}}{\kappa_{d}}\int_{t_{r}}^{\infty}dt\frac{d}{dt}\left(|c_{10}|^{2}+|c_{01}|^{2}\right)\nonumber \\
 & = & \frac{\kappa_{d,1}}{\kappa_{d,0}+\kappa_{d,1}}\left(|c_{10}(t_{r})|^{2}+|c_{01}(t_{r})|^{2}-|c_{10}(\infty)|^{2}-|c_{01}(\infty)|^{2}\right).
\end{eqnarray}
If there is no excitation in the cavity after the retrieval process,
$|c_{10}(\infty)|^{2}=|c_{01}(\infty)|^{2}=0$, the efficiency reduces
to
\begin{equation}
\eta_{s}=\frac{\kappa_{d,1}}{\kappa_{d,0}+\kappa_{d,1}}.
\end{equation}
Therefore, the retrieval efficiency is independent of the pulse shape
of the driving fields, provided that the driving field pumps all excitation
out of the cavity completely. To achieve high retrieval efficiency,
over-coupled condition of the waveguide-cavity system is required.

In the following, we derive the relationship between the pulse shapes
of the driving field and the output field. For adiabatic retrieval
process, the input and driving fields are smooth. It is reasonable
to adiabatically eliminate $c_{10}$, the population of state $|10\rangle$,
in the coupled equations, if $\kappa_{d}\gg g/T$. Then Eq.~(\ref{eq:cme1-1})
transforms to
\begin{eqnarray}
-\kappa_{d}c_{10}-i\Omega c_{01} & = & 0.
\end{eqnarray}
Combining it with Eq.~(\ref{eq:cme1-2}), we get 
\begin{eqnarray}
\frac{d}{dt}c_{01} & = & -\frac{|\Omega(t)|^{2}}{\kappa_{d}}c_{01}.
\end{eqnarray}
Solving this differential equation and using the boundary condition
$c_{01}(0)=1$ (the start time of the retrieval process is set to
be $0$ for simplicity),
\begin{eqnarray}
c_{01}(t) & = & e^{-\frac{1}{\kappa_{d}}\int_{0}^{t}|g(t')|^{2}dt'}.
\end{eqnarray}
Denote $h(t,t')=\int_{t}^{t'}|\Omega(t^{''})|^{2}dt^{''}$ and submit
is into Eq.~(\ref{eq:cme1-3}), the output field is derived as
\begin{eqnarray}
\phi_{out}(t) & = & i\frac{\sqrt{2\kappa_{d.1}}}{\kappa_{d}}\Omega(t)e^{-\frac{1}{\kappa_{d}}h(0,t)}.\label{eq:out}
\end{eqnarray}
Computing the retrieval efficiency using Eq.~(\ref{eq:out}),
\begin{eqnarray}
\eta_{r} & = & \int_{0}^{\infty}dt|\phi_{out}|^{2}\nonumber \\
 & = & \frac{2\kappa_{d,1}}{\kappa_{d}^{2}}\int_{0}^{\infty}dt|\Omega(t)|^{2}e^{-\frac{2}{\kappa_{d}}\int_{0}^{t}dt'|g(t')|^{2}}\nonumber \\
 & = & \frac{\kappa_{d,1}}{\kappa_{d}}\left(e^{-\frac{2}{\kappa_{d}}h(0,0)}-e^{-\frac{2}{\kappa_{d}}h(0,\infty)}\right)\nonumber \\
 & = & \frac{\kappa_{d,1}}{\kappa_{d,0}+\kappa_{d,1}}\left(1-e^{-\frac{2}{\kappa_{d}}h(0,\infty)}\right).\label{eq:out-prac}
\end{eqnarray}
For strong or long driving pulse, $h(0,\infty)\gg1,$\textbf{ }the
retrieval efficiency reduces to the previously derived form $\frac{\kappa_{d,1}}{\kappa_{d,0}+\kappa_{d,1}}$.

To be more practical, we consider the case that not all excitation
are driven out of the cavity, which means either $c_{01}(\infty)$
or $c_{10}(\infty)$ does not equal to zero. The efficiency can be
obtained by giving $h(0,\infty)$ an finite value. Here, we shape
the driving field $\Omega(t)$ to retrieve the photon to the desired
mode $A_{in}(t)$ (the input shpae). Equaling $\sqrt{\frac{\kappa_{d,1}}{\kappa_{d,0}+\kappa_{d,1}}}A_{in}(t)$
and eq.\ref{eq:out} and their norm square integrals,
\begin{eqnarray}
\sqrt{\frac{\kappa_{d,1}}{\kappa_{d,0}+\kappa_{d,1}}}A_{in}(t) & = & i\frac{\sqrt{2\kappa_{d.1}}}{\kappa_{d}}\Omega(t)e^{-\frac{1}{\kappa_{d}}h(0,t)},
\end{eqnarray}
so
\begin{eqnarray}
\int_{0}^{t}dt'|A_{in}(t)|^{2} & = & 1-e^{-\frac{2}{\kappa_{d}}h(0,t)}.
\end{eqnarray}
We derive the optimal retrieval pulse shape
\begin{eqnarray}
\Omega(t) & = & -i\sqrt{\frac{\kappa_{d}}{2}}e^{h(0,t)/\kappa_{c}}A_{in}(t)\nonumber \\
 & = & -i\sqrt{\frac{\kappa_{d}}{2}}\frac{A_{in}(t)}{\sqrt{\int_{t}^{\infty}dt'|A_{in}(t')|^{2}}}.
\end{eqnarray}
For photon storage, it is the time reverse of the retrieve process
\cite{storage}. One can use driving pulse $\Omega^{*}(t_{r}-t)$
to store input pulse of shape $A_{in}\left(t\right)$ with maximal
efficiency.

\subsection{Transmission of the target photon}

The target photon ($a_{in}\left(t\right)$) can also be coupled to
the artificial atom mediated by the antenna under an EO drive. However,
when the two modes $a$ and $b$ in the artificial atom are designed
to fulfill phase-match condition with mode $c$, whether the target
photon can enter into mode $b$ of the artificial atom ($|1_{a}1_{b}\rangle$)
strongly depends on the quantum state of mode $a$. For sufficiently
large nonlinear coupling strength $g_{\mathrm{nd}}$, the degeneracy
of two-excitation states $|1_{a}1_{b}0_{c}\rangle$ and $|0_{a}0_{b}1_{c}\rangle$
will be removed, resulting in two new eigenstates with the form of
superposition of $|1_{a}1_{b}0_{c}\rangle$ and $|0_{a}0_{b}1_{c}\rangle$
and eigenenergiessplited by $2g_{\mathrm{nd}}$. The sum of the two
photon frequencies no longer equals to the resonant frequency of two-excitation
states. If the control photon has already been stored into mode $b$,
the target photon is no longer resonant with the transition from $|0_{a}1_{b}0_{c}\rangle$
to $|1_{a}1_{b}0_{c}\rangle$, thus will be blocked. Accordingly,
the phase of the target photon is controlled by the state of the stored
control photon in the artificial atom.

In this work, we take the artificial atom based on non-degenerate
three-wave mixing for constructing quantum phase gate, thus the total
system Hamiltonian is 
\begin{eqnarray}
H & = & \omega_{\mathrm{d}}d^{\dagger}d+\omega_{\mathrm{a}}a^{\dagger}a+\omega_{\mathrm{a}}b^{\dagger}b+\delta_{c}c^{\dagger}c+\Omega a^{\dagger}d+g_{nd}abc^{\dagger}+h.c.
\end{eqnarray}
 For EO driving strength $\kappa_{A}\ll\Omega\ll g_{nd}$, the waveguide
is effectively over-coupled with the atom. The target photon couples
into and out from the atom for the control photon in state $|0\rangle$,
whereas transmit directly without entering the atom for control photon
in state $|1\rangle$. In this case, it is no longer necessary to
design the shape of the EO drive, but requires carefully choosing
an appropriate strength.

In the two-excitation subspace, the state can be described by a pure
state with a general form of the superposition of Fock states. For
mode $a$ in state $|1\rangle$ and very long lifetime, the single-photon
state is expressed as
\begin{eqnarray}
|\psi\rangle_{dbac} & = & c_{1010}|1010\rangle+c_{0110}|0110\rangle+c_{0001}|0001\rangle,
\end{eqnarray}
where $|mnkl\rangle=|m\rangle_{d}\otimes|n\rangle_{a}\otimes|k\rangle_{b}\otimes|l\rangle_{c}$.
The dynamics follows
\begin{eqnarray}
\frac{d}{dt}c_{mnkl} & = & \frac{d}{dt}\langle mnkl|\psi\rangle\nonumber \\
 & = & \langle mnkl|\frac{d}{dt}\psi\rangle\nonumber \\
 & = & -i\langle mnkl|H_{eff}|\psi\rangle\nonumber \\
 & = & \alpha_{d}\sum_{m'n'k'l'}c_{m'n'k'l'}\langle mnkl|d^{\dagger}d|m'n'k'l'\rangle+\alpha_{b}\sum_{m'n'k'l'}c_{m'n'k'l'}\langle mnkl|b^{\dagger}b|m'n'k'l'\rangle\nonumber \\
 &  & +\alpha_{a}\sum_{m'n'k'l'}c_{m'n'k'l'}\langle mnkl|a^{\dagger}a|m'n'k'l'\rangle+\alpha_{c}\sum_{m'n'k'l'}c_{m'n'k'l'}\langle mnkl|c^{\dagger}c|m'n'k'l'\rangle\nonumber \\
 &  & -i\Omega\left(\sum_{m'n'k'l'}c_{m'n'k'l'}\langle mnkl|a^{\dagger}d|m'n'k'l'\rangle+\sum_{m'n'k'l'}c_{m'n'k'l'}\langle mnkl|ad^{\dagger}|m'n'k'l'\rangle\right)\nonumber \\
 &  & -ig_{nd}\left(\sum_{m'n'k'l'}c_{m'n'k'l'}\langle mnkl|a^{\dagger}b^{\dagger}c|m'n'k'l'\rangle+\sum_{m'n'k'l'}c_{m'n'k'l'}\langle mnkl|abc^{\dagger}|m'n'k'l'\rangle\right)\nonumber \\
 & = & \alpha_{d}mc_{mnkl}+\alpha_{a}nc_{mnkl}+\alpha_{b}kc_{mnkl}+\alpha_{c}lc_{mnkl}\nonumber \\
 &  & -i\Omega\sqrt{m(n+1)}c_{(m-1)(n+1)kl}-i\Omega\sqrt{(m+1)n}c_{(m+1)(n-1)kl}\nonumber \\
 &  & -ig_{nd}\sqrt{nk(l+1)}c_{m(n-1)(k-1)(l+1)}-ig_{nd}\sqrt{(n+1)(k+1)l}c_{m(n+1)(k+1)(l-1)}.
\end{eqnarray}
For our two-photon case
\begin{eqnarray}
\frac{d}{dt}c_{1010} & = & \alpha_{d}c_{1010}+\alpha_{b}c_{1010}-i\Omega c_{0110}-i\phi_{in}(t),\label{eq:cme2-1}\\
\frac{d}{dt}c_{0110} & = & \alpha_{b}c_{0110}+\alpha_{a}c_{0110}-i\Omega c_{1010}-ig_{nd}c_{0001},\label{eq:cme2-2}\\
\frac{d}{dt}c_{0001} & = & \alpha_{c}c_{0001}-ig_{nd}c_{0110}.\label{eq:cme2-3}
\end{eqnarray}
The dynamics of the system depends on the initial state of the system.
Initially, we set $c_{1010}=c_{0110}=c_{0001}=0$ to solve the dynamics
of the system.
\end{document}